\input harvmac
\input epsf
\noblackbox
\overfullrule=0pt
\def\Title#1#2{\rightline{#1}\ifx\answ\bigans\nopagenumbers\pageno0\vskip1in
\else\pageno1\vskip.8in\fi \centerline{\titlefont #2}\vskip .5in}
 
scaled\magstep3 
 
scaled\magstep3 
 
scaled\magstep3 
 
\font\cmss=cmss10 \font\cmsss=cmss10 at 7pt
\def\ove#1#2{\vbox{\halign{$##$\cr
#1{\leftarrow}\mkern-6mu\cleaders\hbox{$#1\mkern-2mu{-}\mkern-2mu$}\hfill
       \mkern-12mu{\to}\cr
    \noalign{\kern -1pt\nointerlineskip}
      \hfil#1#2\hfil\cr}}}
\def\ov#1{\ove{}#1}
%

\newcount\figno
\figno=0
\def\fig#1#2#3{
\par\begingroup\parindent=0pt\leftskip=1cm\rightskip=1cm\parindent=0pt
\baselineskip=11pt \global\advance\figno by 1 \midinsert
\epsfxsize=#3 \centerline{\epsfbox{#2}} \vskip 12pt {\bf Fig.\
\the\figno: } #1\par
\endinsert\endgroup\par
}
\def\figlabel#1{\xdef#1{\the\figno}}
\def\encadremath#1{\vbox{\hrule\hbox{\vrule\kern8pt\vbox{\kern8pt
\hbox{$\displaystyle #1$}\kern8pt} \kern8pt\vrule}\hrule}}
\font\cmss=cmss10 \font\cmsss=cmss10 at 7pt

\def\IB{\relax\hbox{$\inbar\kern-.3em{\rm B}$}}
\def\IC{\relax\hbox{$\inbar\kern-.3em{\rm C}$}}
\def\IQ{\relax\hbox{$\inbar\kern-.3em{\rm Q}$}}
\def\ID{\relax\hbox{$\inbar\kern-.3em{\rm D}$}}
\def\IE{\relax\hbox{$\inbar\kern-.3em{\rm E}$}}
\def\IF{\relax\hbox{$\inbar\kern-.3em{\rm F}$}}
\def\IG{\relax\hbox{$\inbar\kern-.3em{\rm G}$}}
\def\IGa{\relax\hbox{${\rm I}\kern-.18em\Gamma$}}
\def\IH{\relax{\rm I\kern-.18em H}}
\def\IK{\relax{\rm I\kern-.18em K}}
\def\IL{\relax{\rm I\kern-.18em L}}
\def\IP{\relax{\rm I\kern-.18em P}}
\def\IR{\relax{\rm I\kern-.18em R}}
\def\Z{\relax\ifmmode\mathchoice
{\hbox{\cmss Z\kern-.4em Z}}{\hbox{\cmss Z\kern-.4em Z}}
{\lower.9pt\hbox{\cmsss Z\kern-.4em Z}} {\lower1.2pt\hbox{\cmsss
Z\kern-.4em Z}}\else{\cmss Z\kern-.4em Z}\fi}

\def\II{\relax{\rm I\kern-.18em I}}

\def\p{\partial}

\lref\DiFrancescoSS{
P.~Di Francesco and D.~Kutasov,
Phys.\ Lett.\ B {\bf 261}, 385 (1991).
}

\lref\DiFrancescoUD{
P.~Di Francesco and D.~Kutasov,
Nucl.\ Phys.\ B {\bf 375}, 119 (1992)
[arXiv:hep-th/9109005].
}

\lref\GinspargIS{
P.~H.~Ginsparg and G.~W.~Moore,
arXiv:hep-th/9304011.
}

\lref\SchomerusVV{ V.~Schomerus, ``Rolling tachyons from Liouville
theory,'' arXiv:hep-th/0306026.
}
\lref\ftt{M.~Headrick, S.~Minwalla and T.~Takayanagi, ``Closed string
tachyon condensation: an overview'', to be published.}
\lref\MooreZV{
G.~W.~Moore, M.~R.~Plesser and S.~Ramgoolam,
``Exact S matrix for 2-D string theory,''
Nucl.\ Phys.\ B {\bf 377}, 143 (1992)
[arXiv:hep-th/9111035].
}
\lref\greg{
G.~W.~Moore,
``Double scaled field theory at c = 1,''
Nucl.\ Phys.\ B {\bf 368}, 557 (1992).
}

\lref\WittenYR{
E.~Witten,
``On string theory and black holes,''
Phys.\ Rev.\ D {\bf 44}, 314 (1991).
}
\lref\TakayanagiSM{ T.~Takayanagi and N.~Toumbas, ``A matrix model
dual of type 0B string theory in two dimensions,'' JHEP {\bf
0307}, 064 (2003) [arXiv:hep-th/0307083].
}

\lref\DaCunhaFM{
B.~C.~Da Cunha and E.~J.~Martinec,
Phys.\ Rev.\ D {\bf 68}, 063502 (2003)
[arXiv:hep-th/0303087].
}

\lref\GrossAY{ D.~J.~Gross and N.~Miljkovic, ``A Nonperturbative
Solution of $D = 1$ String Theory,'' Phys.\ Lett.\ B {\bf 238},
217 (1990);
}
%
\lref\BrezinSS{ E.~Brezin, V.~A.~Kazakov and A.~B.~Zamolodchikov,
``Scaling Violation in a Field Theory of Closed Strings in One
Physical Dimension,'' Nucl.\ Phys.\ B {\bf 338}, 673 (1990);
}
%
\lref\GinspargAS{ P.~Ginsparg and J.~Zinn-Justin, ``2-D Gravity +
1-D Matter,'' Phys.\ Lett.\ B {\bf 240}, 333 (1990).
}
\lref\DasKA{ S.~R.~Das and A.~Jevicki, ``String Field Theory And
Physical Interpretation Of D = 1 Strings,'' Mod.\ Phys.\ Lett.\ A
{\bf 5}, 1639 (1990).
}
\lref\DouglasUP{ M.~R.~Douglas, I.~R.~Klebanov, D.~Kutasov,
J.~Maldacena, E.~Martinec and N.~Seiberg, ``A new hat for the c =
1 matrix model,'' arXiv:hep-th/0307195.
}
\lref\StromingerFN{ A.~Strominger and T.~Takayanagi, ``Correlators
in timelike bulk Liouville theory,'' arXiv:hep-th/0303221.
}

\lref\kkk{ V.~Kazakov, I.~K.~Kostov and D.~Kutasov, `A matrix
model for the two-dimensional black hole,'' Nucl.\ Phys.\ B {\bf
622}, 141 (2002) [arXiv:hep-th/0101011].
}
\lref\GrossUB{ D.~J.~Gross and I.~R.~Klebanov, ``One-Dimensional
String Theory On A Circle,'' Nucl.\ Phys.\ B {\bf 344}, 475
(1990).
}
\lref\sen{ A.~Sen, ``Rolling tachyon,'' JHEP {\bf 0204}, 048
(2002) [arXiv:hep-th/0203211].
}
\lref\mgas{ M.~Gutperle and A.~Strominger, ``Spacelike branes,''
JHEP {\bf 0204}, 018 (2002) [arXiv:hep-th/0202210].
}
\lref\MinicRK{ D.~Minic, J.~Polchinski and Z.~Yang, ``Translation
Invariant Backgrounds In (1+1)-Dimensional String Theory,'' Nucl.\
Phys.\ B {\bf 369}, 324 (1992).
}
\lref\MartinecKA{ E.~J.~Martinec, ``The annular report on
non-critical string theory,'' arXiv:hep-th/0305148.
}
\lref\PolchinskiMB{
J.~Polchinski,
``What is string theory?,''
arXiv:hep-th/9411028.
}

\lref\kms{ I.~R.~Klebanov, J.~Maldacena and N.~Seiberg, ``D-brane
decay in two-dimensional string theory,'' JHEP {\bf 0307}, 045
(2003) [arXiv:hep-th/0305159].
}
\lref\GinspargIS{
P.~Ginsparg and G.~W.~Moore,
``Lectures On 2-D Gravity And 2-D String Theory,''
arXiv:hep-th/9304011.
}
\lref\hv{ J.~McGreevy and H.~Verlinde, ``Strings from tachyons:
The c = 1 matrix reloaded,'' arXiv:hep-th/0304224.
}
\lref\msy{ A.~Maloney, A.~Strominger and X.~Yin, ``S-brane
thermodynamics,'' arXiv:hep-th/0302146.
}
\lref\BrezinRB{ E.~Brezin and V.~A.~Kazakov, ``Exactly Solvable
Field Theories Of Closed Strings,'' Phys.\ Lett.\ B {\bf 236}, 144
(1990).
} 

\lref\bd{
N.~D.~Birrell and P.~C.~W.~Davies,
``Quantum Fields In Curved Space,''
Cambridge Univ. Pr. (1982).
}

\lref\GrossVS{ D.~J.~Gross and A.~A.~Migdal, ``Nonperturbative
Two-Dimensional Quantum Gravity,'' Phys.\ Rev.\ Lett.\  {\bf 64},
127 (1990).
}
\lref\DouglasVE{ M.~R.~Douglas and S.~H.~Shenker, ``Strings In
Less Than One-Dimension,'' Nucl.\ Phys.\ B {\bf 335}, 635 (1990).
}
\lref\PolchinskiUQ{ J.~Polchinski, ``Classical Limit Of
(1+1)-Dimensional String Theory,'' Nucl.\ Phys.\ B {\bf 362}, 125
(1991).
} \lref\KlebanovQA{ I.~R.~Klebanov, ``String theory in
two-dimensions,'' arXiv:hep-th/9108019.
}

\lref\NatsuumeSP{ M.~Natsuume and J.~Polchinski, ``Gravitational
Scattering In The C = 1 Matrix Model,'' Nucl.\ Phys.\ B {\bf 424},
137 (1994) [arXiv:hep-th/9402156].
}

\lref\KarczmarekXM{ J.~L.~Karczmarek, H.~Liu, J.~Maldacena and
A.~Strominger, ``UV finite brane decay,'' arXiv:hep-th/0306132.
}

\lref\GaiottoRM{ D.~Gaiotto, N.~Itzhaki and L.~Rastelli, ``Closed
strings as imaginary D-branes,'' arXiv:hep-th/0304192.
}

\lref\PolchinskiJP{ J.~Polchinski, ``On the nonperturbative
consistency of d = 2 string theory,'' Phys.\ Rev.\ Lett.\  {\bf
74}, 638 (1995) [arXiv:hep-th/9409168].
}

\lref\AlexandrovCM{
S.~Alexandrov and V.~Kazakov,
``Correlators in 2D string theory with vortex condensation,''
Nucl.\ Phys.\ B {\bf 610}, 77 (2001)
[arXiv:hep-th/0104094].
}

\lref\AlexandrovFH{
S.~Y.~Alexandrov, V.~A.~Kazakov and I.~K.~Kostov,
``Time-dependent backgrounds of 2D string theory,''
Nucl.\ Phys.\ B {\bf 640}, 119 (2002)
[arXiv:hep-th/0205079].
}

\lref\AlexandrovPZ{
S.~Y.~Alexandrov and V.~A.~Kazakov,
``Thermodynamics of 2D string theory,''
JHEP {\bf 0301}, 078 (2003)
[arXiv:hep-th/0210251].
}

\lref\AlexandrovQK{
S.~Y.~Alexandrov, V.~A.~Kazakov and I.~K.~Kostov,
``2D string theory as normal matrix model,''
Nucl.\ Phys.\ B {\bf 667}, 90 (2003)
[arXiv:hep-th/0302106].
}
\lref\AlexandrovUH{ S.~Alexandrov, ``Backgrounds of 2D string
theory from matrix model,'' arXiv:hep-th/0303190.
}
\lref\AlexandrovUT{ S.~Alexandrov, ``Matrix quantum mechanics and
two-dimensional string theory in non-trivial backgrounds,''
arXiv:hep-th/0311273.
}
\lref\MinwallaHJ{
S.~Minwalla and T.~Takayanagi,
``Evolution of D-branes under closed string tachyon condensation,''
JHEP {\bf 0309}, 011 (2003)
[arXiv:hep-th/0307248].

}
\lref\DavidVM{
J.~R.~David, M.~Gutperle, M.~Headrick and S.~Minwalla,
``Closed string tachyon condensation on twisted circles,''
JHEP {\bf 0202}, 041 (2002)
[arXiv:hep-th/0111212].
}

\lref\AdamsSV{
A.~Adams, J.~Polchinski and E.~Silverstein,
``Don't panic! Closed string tachyons in ALE space-times,''
JHEP {\bf 0110}, 029 (2001)
[arXiv:hep-th/0108075].
}
\lref\GutperleMB{
M.~Gutperle and A.~Strominger,
``Fluxbranes in string theory,''
JHEP {\bf 0106}, 035 (2001)
[arXiv:hep-th/0104136].
}

\lref\SchomerusVV{
V.~Schomerus,
JHEP {\bf 0311}, 043 (2003)
[arXiv:hep-th/0306026].
}

\lref\SeibergBJ{ N.~Seiberg and S.~H.~Shenker, ``A Note on
background (in)dependence,'' Phys.\ Rev.\ D {\bf 45}, 4581 (1992)
[arXiv:hep-th/9201017].
}
\lref\ShenkerUF{
S.~H.~Shenker,
``The Strength Of Nonperturbative Effects In String Theory,''
RU-90-47
{\it Presented at the Cargese Workshop on Random Surfaces, Quantum Gravity and Strings, Cargese, France, May 28 - Jun 1, 1990}
}
\lref\KarczmarekPV{ J.~L.~Karczmarek and A.~Strominger, ``Matrix
cosmology,'' arXiv:hep-th/0309138.
}
\lref\McGreevyEP{
J.~McGreevy, J.~Teschner and H.~Verlinde,
``Classical and quantum D-branes in 2D string theory,''
JHEP {\bf 0401}, 039 (2004)
[arXiv:hep-th/0305194].
}
\lref\DabholkarWN{
A.~Dabholkar and C.~Vafa,
``tt* geometry and closed string tachyon potential,''
JHEP {\bf 0202}, 008 (2002)
[arXiv:hep-th/0111155].
}
\lref\McGreevyKB{
J.~McGreevy and H.~Verlinde,
``Strings from tachyons: The c = 1 matrix reloaded,''
JHEP {\bf 0312}, 054 (2003)
[arXiv:hep-th/0304224].
}
\lref\cta{
Y.~Okawa and B.~Zwiebach,
``Twisted Tachyon Condensation in Closed String Field Theory,''
arXiv:hep-th/0403051.
}
\lref\ctb{
M.~Headrick,
arXiv:hep-th/0312213.
}
\lref\ctc{
S.~Sarkar and B.~Sathiapalan,
``Closed string tachyons on C/Z(N),''
arXiv:hep-th/0309029.
}
\lref\bta{
T.~Suyama,
``On decay of bulk tachyons,''
arXiv:hep-th/0308030.
}
\lref\btb{
A.~A.~Tseytlin,
``Magnetic backgrounds and tachyonic instabilities in closed string  theory,''
arXiv:hep-th/0108140.
}
%
\Title{\vbox{\baselineskip12pt \hbox{hep-th/0403169}}} {\vbox{
\centerline {Closed String Tachyon Condensation at $c=1$}}}
\centerline{Joanna L. Karczmarek
  and  Andrew Strominger}
\vskip.1in {\it Jefferson Physical Laboratory, Harvard
University, Cambridge, MA 02138}

\vskip.1in \centerline{\bf Abstract} The $c=1$ matrix model, with
or without a type 0 hat, has an exact quantum solution
corresponding to closed string tachyon condensation along a null
surface. The condensation occurs, and spacetime dissolves, at a
finite retarded time on ${\cal I}^+$. The outgoing quantum state
of tachyon fluctuations in this time-dependent background is
computed using both the collective field and exact fermion
pictures. Perturbative particle production induced by the
moving tachyon wall is shown to be similar to that induced by a
soft moving mirror. Hence, despite the fact that ${\cal I}^+$ for
the tachyon is geodesicaly incomplete, quantum correlations in the
incoming state are unitarily transmitted to the outgoing state in
perturbation theory. It is also shown that, non-perturbatively,
information can leak across the tachyon wall, and tachyon
scattering is not unitary. Exact unitarity remains intact only in
the free fermion picture.

\Date{}

\listtoc\writetoc

%
\newsec{Introduction}

The $c= 1$  matrix model  provides a non-perturbatively soluble string
theory in 1+1 dimensions.
However, despite a wealth of detailed data, the physical
interpretation of many aspects of the theory remained unclear,
especially during the first chapter of the story a decade ago. The
hoped-for nonperturbative lessons did not fully materialize (one
notable exception is \ShenkerUF ) and the nonperturbative
consistency of the theory was questioned \PolchinskiJP.
A cogent summary of the lessons and disappointements can be found
in the conclusions of \GinspargIS.

   A new chapter of the $c=1$ story was recently opened with the reinterpretation of the
theory as an example of open-closed holography \refs{\McGreevyKB,
\kms} and as a type 0 string theory\refs{\TakayanagiSM,
\DouglasUP}.  As a result, the study of some non-perturbative
stringy phenomena became possible. Much light was shed on the
problem of unstable D-brane decay \refs{\kms,\McGreevyEP}, which
can be viewed as $open$ string tachyon condensation. A family of
exact solutions to the theory, found in
\refs{\MinicRK,\AlexandrovFH}, were shown to correspond physically
to the time-dependent process of closed string tachyon
condensation\foot{We follow the standard abuse of notation in
referring to the massless field of the $c=1$ theory as a tachyon.
We note that the tachyon cannot condense spontaneously in the
usual vacuum, which is perturbatively stable. The process we
consider involves a change of boundary conditions at infinity.
Nevertheless we hope it provides a useful toy model for higher
dimensional spontaneous tachyon condensation. Since the tachyon
mode is in fact massless, the usual problem with defining initial
state for an unstable mode does not arise.} \KarczmarekPV. Exact
closed cosmologies were also constructed \KarczmarekPV.

The study of $open$ string tachyon condensation, which can be
described by an exact boundary CFT \sen, has led to many insights
into the nonperturbative character of time-dependent string
theory.  Studies of $closed$ string  tachyon condensation have
been more limited (see the review \ftt ). The case of localized
tachyons was considered in \refs{\AdamsSV \DabholkarWN \ctc \ctb
-\cta } and some aspects of bulk tachyons in \refs{\GutperleMB
\SchomerusVV \btb \bta \DavidVM \MinwallaHJ  \DaCunhaFM- \StromingerFN}. The
main problem, of course, is that spacetime disappears altogether
when a bulk tachyon condenses and hence perturbation theory breaks
down. The existence of non-perturbatively exact quantum solutions
at $c=1$  provides an ideal framework to study this process.

This paper considers the  exact quantum description of closed
string  tachyon condensation along a null hypersurface.  In this
process the Fermi sea -- together with the spacetime described by
its fluctuating surface -- comes to an end at a finite retarded
time $t^-_{end}$ on ${\cal I}^+$. The everywhere-timelike tachyon
wall -- which is static in the far past and accelerates up to
$t^-_{end}$ -- distorts the perturbative incoming vacuum much like
a moving mirror softened at the string scale. Perturbatively, this
moving mirror reflects all the information incoming from ${\cal
I}^-$ into an outgoing excited state in the region of  ${\cal
I}^+$ $prior$ to $t^-_{end}$.

The nonperturbative
picture can be understood in the free fermion formulation.
We find that nonperturbatively information/correlations can leak
through the tachyon wall into the no-man's land beyond the end of
time $t^-_{end}$. This means there is no exact unitary S-matrix for tachyon
scattering. In the free fermion picture there is exact unitarity, but it
involves states with no spacetime interpretation.

So, in this example, exact unitarity of the holographic dual does not imply
the existence of an exactly unitary spacetime picture.

In addition to the intrinsic interest of tachyon condensation, another
motivation for the present work is towards an understanding of
perhaps the most tantalizing of all nonperturbative phenomena:
black hole formation/evaporation, in the $c=1$ context.
One possibility is that black holes simply are not part of the
theory, or cannot be made as
   any kind of
excitation of the vacuum. Another is that they do occur but we have
   not yet recognized them. Adding to the mystery and frustration
is the existence of an
exact worldsheet CFT describing the eternal black hole \WittenYR\ as well as a
deformed matrix model describing the euclidean version \kkk.
As in the AdS/CFT correspondence, the observables of the matrix model
are best thought of as living at the spacetime boundary, namely
${\cal I}^\pm$.  The physical interpretation of quantities in the interior
is obscured by non-local transforms and strong coupling. Hence, black hole
formation might best be recognized by the appearance of Hawking radiation
at ${\cal I}^+$. In the present work, the
 groundwork for the understanding
of such quantum particle production is laid.

A spacetime intepretation of the free fermion variables on the past
and future boundaries ${\cal I}^\pm$ involves
both a bosonization and a nonlocal ``leg pole'' transform.
The working assumption of this paper is that this map and the set
of boundary observables are the same for the large processes such
as tachyon condensation considered here as for the static fermi sea. 
This assumption leads to an apparently self-consistent picture. 
However, we believe that it is quite possible that there are other 
interesting interpretations of the boundary data on ${\cal I}^\pm$  
that differ from the one employed herein perhaps by field or coordinate
redefintions, although we have no specific proposal along these
lines.

This paper is organized as follows. In section 2 we review the
semiclassical matrix model solution corresponding to tachyon
condensation along an asymptotically null hypersurface. In the
free fermion picture it corresponds to a draining of the Fermi
sea, while in the spacetime picture the initially static tachyon
wall accelerates up to ${\cal I}^+$. In section 3 the
semiclassical quantum fluctuations around this solution are
described using the Das-Jevicki\DasKA\ collective field formalism,
together with the non-local leg pole transform \PolchinskiMB\
which turns the collective quantum fluctuations of the Fermi
surface into spacetime tachyons.  We find that, prior to the leg
pole transform,  the motion of the tachyon wall deforms the
quantum vacuum like an impenetrable accelerating  mirror which
reaches ${\cal I}^+$ at the end of the world at $t^-_{end}$. The
outgoing state of the quantum field is computed. Perturbatively,
all the correlations of the incoming vacuum state of the
collective field are transmitted to an excited outgoing state with
support prior to  $t^-_{end}$. The collective field formalism
gives no information about the quantum state in the no-man's land
after $t^-_{end}$.  The leg pole transform leads to a similar
picture for the spacetime tachyon, except that the mirror is soft
rather than impenetrable and spacetime ends less abruptly. A
concrete measure of the deviation of the physics on ${\cal I}^+$
from that of an ordinary massless field is the commutator $\Delta$
of the tachyon field with its time derivative. For normalizable
fluctuations of the static Fermi surface this remains precisely a
delta function, but here we find (due to the large nature of the
background fluctuation) an exact expression which decays
exponentially to zero after the end of the world. In section 4 we
redo the collective field analysis using the exact, but less
intuitive, free fermion picture. An exact quantum state\foot{More
precisely, since the fluctuations involved in tachyon condensation
are non-normalizable in the sense of \SeibergBJ, this is best
described by a deformed Hamiltonian.} is given whose semiclassical
limit is the null tachyon condensation. An exact computation of
the commutator $\Delta$ is performed and found, to our surprise,
to agree exactly with the collective field computation on ${\cal
I}^+$.  Hence, the free fermion picture corroborates the collective
field analysis. Finally in section 5 nonperturbative effects are
considered. It is argued that non-perturbative excitations -- such
as D-branes -- can escape across the wall into the no-man's land
and never reappear on ${\cal I}^+$ before the end of the world.
Hence, the spacetime tachyon picture misses nonperturbative degrees
of freedom in an essential manner and cannot be exactly unitary.
Although there is exact unitarity in the free fermion picture, we
expect there is no unitary spacetime description of the process of
tachyon condensation.

We expect similar phenomena, in which both sides of the Fermi sea
are drained, to occur in the type 0 matrix models. It would be
interesting to consider this in detail.

We set $\alpha' = 1$ throughout the paper.

\newsec{A moving tachyon wall solution}

In this section, we will briefly review a semiclassical solution
of the $c=1$ matrix model having an interpretation as tachyon
condensation along a null hypersurface \KarczmarekPV.

As is well known, the matrix model reduces to a system of nonrelativistic,
free fermions in 1 space dimension, 
whose potential in the double scaling limit becomes
simply $-\half x^2$, while the number of fermions is taken to
infinity. The classical limit of this system is the motion of an
incompressible Fermi fluid in phase space, $(x,p)$.  The static
solutions of this problem are Fermi surfaces given by $x^2 - p^2 =
2 \mu$.  Perturbations about this static solution reproduce the
perturbative behavior of the 2D Liouville string. The two branches
of the hyperbola can be treated separately in perturbation theory.
Nonperturbatively they interact via fermion tunneling under the
potential.  The spacetime interpretation of the two coupled Fermi
seas together was given in \refs{\TakayanagiSM,\DouglasUP}: the
symmetric and antisymmetric fluctuations of the entire Fermi sea
correspond to the tachyon field and the RR flux in type 0B string
theory, respectively.

Other, time dependent, solutions can
be considered
\refs{\MinicRK,\AlexandrovFH ,\AlexandrovCM, \AlexandrovPZ, \KarczmarekPV,
 \AlexandrovQK}, and our subject is to understand the spacetime
interpretation of one such solution on the quantum level.

A particular solution of the equations of motion for the Fermi
surface which we will focus on is given by \foot{The discussion
will focus of the right branch of the hyperbola until section 4. For a Type 0B solution we should include the second
filled region at negative $x$, but for the sake of
brevity we do not consider this.} \eqn\moving{ (x + p - 2\lambda
e^t)(x - p) = 2 \mu~,} where $\mu,\lambda $ are arbitrary
non-negative constants. An exact quantum representation in terms
of free fermions will be given in section 4. This is a {\it
moving} hyperbola centered at $(x,p)= (\lambda e^t, \lambda e^t )$
rather than the origin. Introducing a parameter $-\infty<\sigma
<\infty $ along the curve, the right branch of the solution to
\moving\ may be written in the alternate useful form
\eqn\trd{\eqalign{x&= \sqrt{2\mu} \cosh \sigma + \lambda e^t ~, \cr
p&= \sqrt{2\mu} \sinh \sigma +\lambda e^t~ .}}
In section 4 we will give an exact quantum description
 in the free Fermi picture of a state
whose Fermi surface obeys \trd\ in the semiclassical limit.

From \moving, we see that in the infinite past the Fermi sea is
essentially static and filled up to the energy $\mu$ below
the top of the potential. It then decays away (rolls down the
potential) and all fermions move out to $x=\infty$ in the infinite
future. Physically, this corresponds to tachyon condensation along
the asymptotically null hypersurface $t-\ln x=-\ln \lambda$
\KarczmarekPV. Using the dictionary of
\refs{\PolchinskiMB-\PolchinskiUQ} in the asymptotic region of
large $x=e^{q}$, the solution \moving\ corresponds to a tachyon
field given, to leading order at large $X^1$, by
 \eqn\T{ T = \hat \mu  X^1 e^{-2 X^1} + \hat \lambda
e^{-X^1+X^0} ~, } where $\hat \lambda$ and $\hat \mu$ are
proportional to $\lambda$ and $\mu$.\foot{A divergence in the leg
pole factor leads to an infinite renormalization of $\hat \lambda
$ in terms of $\lambda$.} Including this field leads to the closed
string worldsheet interaction \eqn\actionB{ {1 \over {4\pi}} \int
d^2\ \sigma \sqrt{\hat g} \Big\{ 2  X^1 \hat R + \hat \mu X^1
e^{-2 X^1} + \hat \lambda e^{-X^1+X^0}  \Big\}~. } The first
interaction term corresponds to standard Liouville potential
which is a timelike ``wall'' at $X^1 \sim \half \ln \hat \mu$. The
second interaction term (which is a weight (1,1) operator) is also
an exponential potential wall, albeit less steep than the standard
one. This wall moves with time, effectively cutting off the
universe at a time-varying distance. It is located roughly along
the outgoing null trajectory at $X^1- X^0\sim \ln \hat \lambda$.
Hence, at large positive  $X^0$, the wall moves outwards  with
essentially the speed of light. The situation is depicted in
Figure 1.

\fig{
Penrose diagram for lightlike tachyon condensation.
The timelike curve ending on ${\cal I}^+$ is the tachyon wall.
There are no perturbative degrees of freedom propagating
in the shaded region.}
{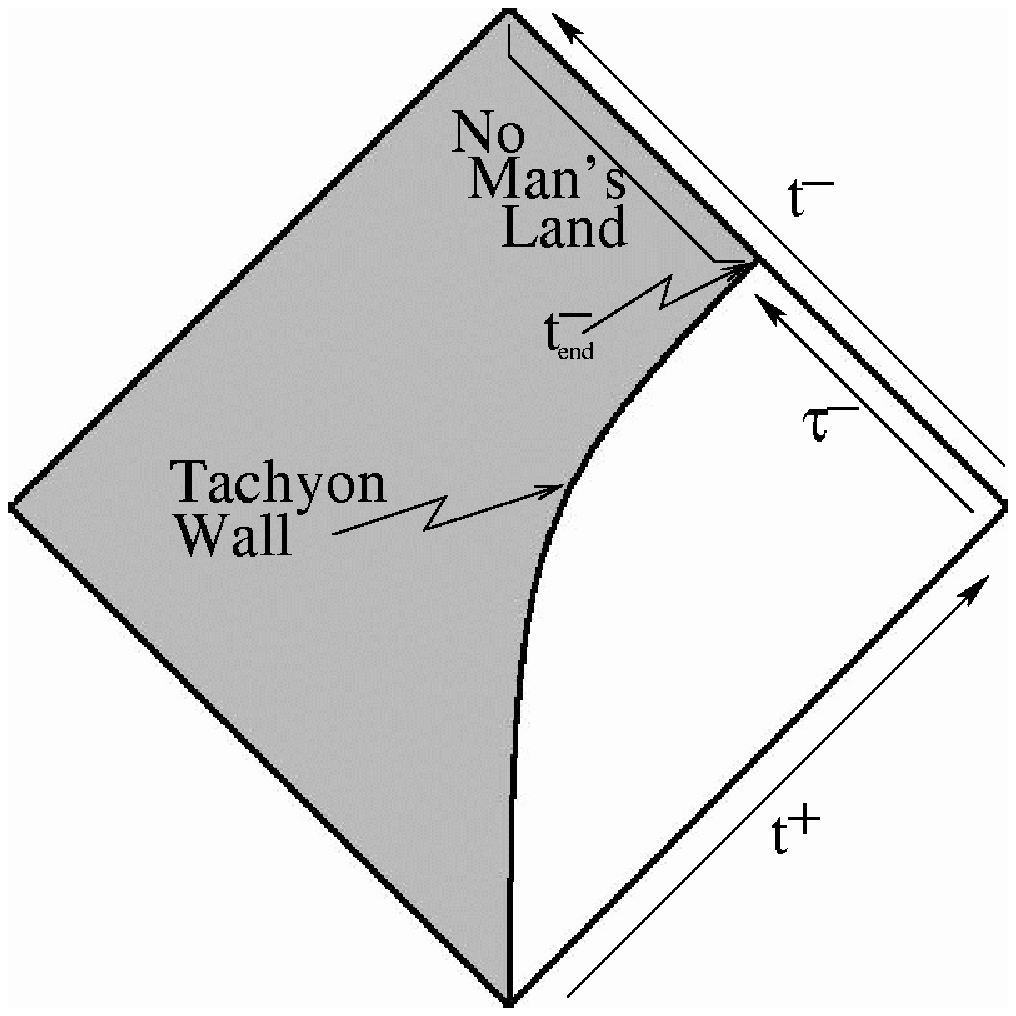}{0.0truein}

 Any observer moving along a timelike
trajectory will eventually move into a region where the tachyon
field becomes arbitrarily large. Therefore this solution is a form
of closed string tachyon condensation \KarczmarekPV.

\newsec{The collective field picture}
The previous section used the dictionary  developed in
\PolchinskiUQ\ to obtain the spacetime interpretation of a
moving Fermi surface.  This dictionary is valid only for small
deviations from the static Fermi surface, and breaks down at late
times for our  solution when the deviations become large. In this
section we will use the (in this respect) more general and
powerful Das-Jevicki collective field formalism \DasKA\ (see also
the recent discussion \AlexandrovUT ). The goal will be to
determine the outgoing quantum state of the tachyon field.

\subsec{Classical action for small fluctuations}
 We are
interested in expressing the effective action for the collective
fermion field. The action, in ``fermion coordinates''
is {\DasKA}  
\eqn\actionxtfull{ \int dt dx
\Big [{Z^2 \over 2\pi\varphi}
-{\varphi^3 \over 6\pi}   + {1 \over \pi}
({1\over 2} x^2 - \mu) \varphi
\Big ]~, } 
where $\varphi$ is defined in terms of the upper and
lower Fermi surfaces $p_\pm$ by \eqn\varphizero{ \varphi(x,t)
\equiv \half (p_+(x,t) - p_-(x,t))~, } while $Z$ is defined by
\eqn\zdef{Z(x,t) \equiv \int^x dx'
\partial_t \varphi(x')~.}
 The collective field $\eta$ describes the difference between a
fluctuating Fermi surface $\varphi$ and a background solution
$\varphi_0$: \eqn\faz{\varphi = \varphi_0 + \sqrt{\pi} \partial_x\eta~.} 
From \actionxtfull, we obtain the action for $\eta$
\eqn\actionxt{ \int dt dx
\Big [ {1\over 2\pi} {(Z_0 + \sqrt{\pi}
\partial_t\eta)^2 \over \varphi_0 + \sqrt{\pi} \partial_x\eta}
-{1\over 6\pi}  (\varphi_0 + \sqrt{\pi}\p_x\eta)^3 + {1 \over \pi}
({1\over 2} x^2 - \mu) (\varphi_0 + \sqrt{\pi} \partial_x \eta)
\Big ]~.} 
In the case of present interest \moving\ we have
\eqn\ijn{\varphi_0=\sqrt{(x-\lambda e^t)^2 - 2 \mu},~~~~Z_0=-\lambda
e^t \varphi_0~.}
The action \actionxt\
contains the quadratic piece
 \eqn\Stwo{
S_{(2)} = \half \int {dt ~dx \over \varphi_0} \Big [ (\partial_t
\eta)^2 - 2 {Z_0\over \varphi_0} (\partial_t \eta
\partial_x \eta) -(\varphi_0^2 - {Z_0^2 \over \varphi_0^2}) (\partial_x
\eta)^2 \Big ]~, } and a leading interaction piece \eqn\Sthree{
S_{(3)} = \int dt ~dx ~ \varphi_0^{-2} o( \eta^3)~. } It is
sometimes  convenient to introduce light cone fermion coordinates
\eqn\erfl{t^\pm= t\pm \ln x =t\pm q~,} in which the quadratic action
\Stwo\ for the solution \ijn\ becomes \eqn\mkg{S_2=\int {dt^+ dt^-
\over \sqrt{(1-\lambda e^{t^-})^2 -2 \mu e^{t^--t^+} }} \bigl[
(1-\lambda e^{t^-})
\p_+\eta\p_-\eta + \lambda e^{t^-} (\p_+\eta)^2 +{\mu \over 2}
e^{t^--t^+}(\p_+\eta-\p_- \eta)^2 \bigr]~. } Note that the last
$\mu$-dependent term vanishes both on ${\cal I}^+$ ($t^+ \to
+\infty,~~t^-$ finite) and ${\cal I}^-$ ($t^- \to -\infty,~~t^+$
finite). However, it plays a crucial role in keeping the tachyons
out of the strong coupling region. On ${\cal I}^-$ the $\lambda$
terms also vanish and $\eta$ is a canonical free boson. The
$\lambda$ terms do not vanish on ${\cal I}^+$, where it
represents the effects of the background null tachyon.

 Following
\AlexandrovUH, it is useful to define the ``Alexandrov coordinates''
$\tau^\pm=\tau\pm \sigma$, with $\sigma$ defined by \trd\ and
$\tau=t$. Specifically, \eqn\coordinates {x =
\sqrt{2 \mu}\cosh(\sigma) + \lambda e^\tau~. } This implies
\eqn\lightcone {t^{\pm} = \tau^\pm \pm  \ln \Big (
{\sqrt{ \mu\over 2}} + \lambda e^{\tau^-} + {\sqrt{\mu \over 2}}
e^{\tau^- -~ \tau^+} \Big ) ~.} It is then easy to show that the
partial derivatives of the Alexandrov coordinates with respect to
the original fermion coordinates can be written as \eqn\partials
{\partial_x \sigma = {1 \over \varphi_0} ~,~~~
\partial_t \sigma = {Z_0 \over \varphi_0^2} ~,~~~
\partial_x \tau = 0 ~,~~~
\partial_t \tau = 1~.
} Using the above set of partial derivatives, it is not hard to
show that the action given in \actionxt\ becomes in the
Alexandrov coordinates \eqn\actiontausigma{
\eqalign{ \int d\tau d\sigma
\Big [ {1\over 2} ((\partial_\tau\eta)^2 - (\partial_\sigma
\eta)^2) &- { \sqrt{\pi} \over 6 \varphi_0^2}
(3(\partial_\tau\eta)^2(\partial_\sigma \eta) + (\partial_\sigma
\eta)^3) \cr &+ \sum_{n=2}^{\infty} {(-1)^n\over 2}(\partial_\tau
\eta)^2
 \Big({\sqrt{\pi} (\partial_\sigma \eta) \over \varphi_0^2}\Big)^n
\Big ]~. }}
 In the $(\tau,\sigma)$ coordinates, $\varphi_0 =
\sqrt{2 \mu} \sinh(\sigma)$, so the above action is static.  In fact,
all effects due to $\lambda \neq 0$ have disappeared, and
\actiontausigma\ is identical to the effective action for the pure
Liouville background $\lambda = 0$. 
In particular, the coupling varies exponentially with the coordinate
$\sigma$: $g \sim \exp(-2\sigma)$, and is independent of time $\tau$. 
The quadratic term for $\eta$
is a canonical free field.  In addition there is a reflecting
boundary condition along the timelike line $\tau^+=\tau^-$ \DasKA,
\eqn\rfl{[\p_+\eta-\p_-\eta]_{\tau^+=\tau^-}=0~.} This line
parameterizes the end of the Fermi sea, where incoming
fluctuations round the bend and turn into outgoing fluctuations.
Now we see the advantage of these coordinates. The general
solution of the linearized wave equation following from \actionxt\
can now easily be found by transforming to Alexandrov coordinates
where they are simply plane waves reflected off the origin.

\subsec{The leg pole transform} The S-matrix for $\eta$ scattering
agrees with that of spacetime tachyon scattering only after
performing the nonlocal leg pole transform 
\refs{\DiFrancescoSS, \DiFrancescoUD, \PolchinskiMB}. We
will now use the leg pole transform
 to convert the effective variable $\eta$ into a dressed
spacetime tachyon $S$.  The leg pole transform is only known
in the asymptotic regimes ${\cal I}^\pm$, and we will focus on
${\cal I}^+$ (since behavior at ${\cal I}^-$ is standard). 
Outgoing waves in this region obey $\eta(\sigma,
\tau) = \eta(\tau-\sigma)$. On ${\cal I}^+$,  $\eta$ will be 
written as a function
of {\it one} variable, and we will always mean it as a function of
$\tau^- = \tau-\sigma$.
Comparing the definition of $\eta$ in \faz\ with the
definition of $\bar S$ in \PolchinskiMB, \eqn\barS{ \varphi =
\half (p_+ - p_-) =
 e^q +\sqrt{\pi}e^{-q} \partial_q \bar S(q,t)~,  ~q\equiv \ln x~ ,
}  we see that \eqn\pol{\bar S = \eta +\pi^{-1/2} \int (x-\varphi_0)~.} 
A point on ${\cal I}^+$ labelled by $t^-=t-q$ corresponds in
the Alexandrov coordinates to \eqn\tausigma 
{\tau^- = t^-  +
\ln(\sqrt{\mu/2}) -\ln\Big(1- {\lambda } e^{t^-} \Big)~.} 
Notice
that $\tau^-$ is a function of $t^-$ (and not $t^+$) only in this
limit. The leg pole transform of $\eta(\tau^-)$ reads
therefore\foot{We note that the leg pole transform is defined
in the $t\pm q$ coordinates on $\cal I^+$.
This is our working definition of the spacetime theory.   In principle
there may be other possibilities.} 
\eqn\legpole{\eqalign{ S(t^-) &= \int dv^- K(v^-) \eta\big( t^- -
v^- +\ln (\sqrt{\mu/2})-
\ln (1- \lambda e^{t^- -v^-} )\big) \cr &= \int
dw^- {K\Big( w^- + \ln\big(1 + \lambda e^{t^- -w^-} \big) \Big)
\over 1 + \lambda e^{t^- -w^-} } \eta(t^- -w^-
+ \ln(\sqrt{\mu/2}))~. }} It is not
difficult to convince oneself that for $\lambda e^{t^- - w^-} \ll 1$,
\eqn\Kapprox {{K\Big( w^- + \ln\big(1 + \lambda e^{t^- -w^- } \big)
\Big) \over 1 + \lambda e^{t^- -w^-} } \approx K(w^-)~.} Hence, the
tachyon $S(t^-)$ is appreciably affected by $\lambda \neq 0$ only in
the region where $t^- - w^-> -\ln \lambda$.

 The kernel $K(v)$ is defined as \PolchinskiMB
\eqn\K {\eqalign {K(v) = &\int_{-\infty}^{\infty} {d\omega \over 2\pi} 
e^{i\omega v}
\Big
(\pi / 2 \Big )^{-i\omega/4} {\Gamma(-i\omega) \over
\Gamma(i\omega)} ~=~ - {z \over 2} J_1 (z) ~=~ {d \over dv}
J_0(z)~,\cr &z(v)\equiv 2 (2/\pi)^{1/8} e^{v/2} .}}
 The asymptotic
behavior is \eqn\Kassympt {K(v) \sim e^v~,~~~~~ v \rightarrow
-\infty } and \eqn\Kassympt {K(v) \sim e^{v/4} \cos(z + \pi/4)
~,~~~~~ v \rightarrow +\infty ~.} $K$ decays exponentially for $v
\rightarrow -\infty$ and grows while oscillating wildly for $v
\rightarrow +\infty$.

We are interested in finding the leg pole transform $S_k(t^-)$ of
an  outgoing mode of $\eta$ of momentum k \eqn\wwr{\eta_k(\tau^-) =
{e^{-ik\tau^- }\over \sqrt{2|k|} }~.} Let's consider first the region
$t^- < -\ln \lambda$, where \Kapprox\ is valid.  Thus, the main
contribution to $S_k(t^-)$ comes from region around $w^- = 0$. As a
result, $S_k(t^-)$ is essentially $\eta_k(\tau^- = t^-
+\ln(\sqrt{\mu/2}))$
 with some smearing.

In order to study the region $t^- > -\ln\lambda$, it turns out to
be more convenient to rewrite \legpole\ (integrating by parts)  as
\eqn\trans { S(t^-) = \int dv^- J_0\Big (z\big(v^- + \ln(1 +
\lambda e^{t^- -v^-}) \big)\Big)
\partial \eta(t^- -v^- + \ln(\sqrt{\mu/2}))~.}
Using the integral representation of $J_0$ \eqn\BesselJ{ J_0(z(y))
= \int {d\omega \over 2\pi} \Big ({\pi \over 2} \Big
)^{-i\omega/4} {\Gamma(-i\omega) \over \Gamma(i\omega+1)}
e^{i\omega y}~, } together with a contour prescription where the
integral over the real line passes {\it over} the pole at
$\omega=0$, we obtain that the leg pole transform of $\eta_k$ is
\eqn\stepone{ S_k(t^-) = \int {d\omega dv^- \over 2 \pi} \Big
({\pi \over 2} \Big )^{-i\omega/4} {\Gamma(-i\omega) \over
\Gamma(i\omega+1)}~ e^{i\omega v^-} ~(1+\lambda e^{t^-
-v^-})^{i\omega}~ \big( {-i k e^{ -ik(t^- -v^- +\ln(\sqrt{\mu/2})
)} \over \sqrt{2|k|}}
\big)~. } 
It is easiest to evaluate the integral over $v^-$ first, 
after taking $\omega$ and $k$ slightly off the real axis to
regularize it. This way, we obtain
\eqn\bogolubov{ S_k(t^-) = \int {d\omega \over 2 \pi} \Big ({\pi
\over 2} \Big )^{-i\omega/4} \lambda^{i(\omega+k)}
{\Gamma(-ik-i\omega) ~\Gamma(1+ik) \over \Gamma(i\omega+1)}
{e^{i\omega(t^-)} \over \sqrt{2|k|}} 
\Big({\sqrt{\mu \over 2}}\Big)^{-ik}~. } This integral can be
evaluated by closing the contour in the lower half plane, and
summing over all the poles of $\Gamma(ik-i\omega)$ (including the
pole at $\omega=k$, according to the contour prescription arising
from regularization of the $v^-$ integral).
The resulting series sums to \eqn\mode {S_k(t^-) = {1
\over \sqrt {2|k|}} \Big({\sqrt {2\mu} z \over \lambda}\Big)^{-ik} J_{-ik}(z)
\Gamma(1+ik)~, } where we have defined \eqn\u {z \equiv 2
(2/\pi)^{1/8}(\lambda e^{t^-})^{1/2}~. }

The dressed tachyon $S(q, t)$ is related to the undressed tachyon
field by $T = e^{\Phi} S$. 
However, we are actually more interested in the
field $S$ itself, which is the properly normalized massless
excitation of the tachyon field.

While it is not possible to extract the dilaton from the data
presented here, the asymptotic form of the dilaton should be
$\Phi \sim -2q$, consistent with the following observation: up to
the smearing present in the transform \legpole, and for $t^- <
-\ln \lambda$, we have  $\sigma = q - \ln (\sqrt{2\mu})$.  The
action \actiontausigma\ has a coupling constant $g \sim \phi_0^{-2} 
\sim e^{-2\sigma}$. This suggest that the dilaton is linear
in the region in which perturbations propagate freely
(i.e. for $t^- << -\ln(\lambda)$).

\subsec{The quantum vacuum for the collective field }

Action \mkg\ is time-dependent. However, in the far past the
background is static and there is a natural incoming  vacuum
state. To determine the outgoing state as usual we must solve the
time-dependent wave equation. In this subsection we
discuss the situation in the fermion light cone
coordinates $t^\pm=t\pm q$ and in the next subsection we
consider the effects of the leg pole transform.

The exact linearized wave equation for $\eta$ is obtained by
varying \mkg:\eqn\izyhNEW{\bigl((1-\lambda e^{t^-}) \p_+\p_- +\lambda
e^{t^-}\p_+\p_++ 
{\mu \over 2}e^{t^- - t^+} ((\p_+-\p_-)^2-(\p_+-\p_-)) 
\bigr)\eta=0~.}  In the far past, on ${\cal I}^-$, the equation
reduces simply to \eqn\iyh{\p_+\p_-\eta=0~,~~~~~~~{\rm{on}}~~ {\cal
I}^-~.} In the far future we have 
\eqn\izyh{\bigl(\p_+\p_- +{\lambda e^{t^-}\over (1-\lambda e^{t^-})}\p_+\p_+
\bigr)\eta+=0~,~~~~~~~\rm{on} ~~ {\cal I}^+~.} 
Notice that outgoing plane
waves, which are functions of $t^-$ only, remain solutions of
\izyh\ on ${\cal I}^+$ for nonzero $\lambda$.
 We seek
solutions of the wave equation which are purely positive frequency
on ${\cal I}^-$ \eqn\pfr{u^{in}_\omega (t^+,t^-) \to {e^{-i\omega
t^+} \over \sqrt{2 \omega}}~,~~~~{\rm{for}} ~~t^-\to -\infty, \omega > 0~.} 
Such solutions are orthonormal \eqn\dio{\langle
u^{in}_{\omega'}|u^{in}_{\omega}\rangle \equiv i\int_{\Sigma}
{d\Sigma^\mu \over 2\pi} u^{in*}_{\omega} \ov{ {\p_\mu}}
u^{in}_{\omega'}=\delta(\omega -\omega')~,} where $\Sigma$ is any
complete spacelike or null slice and $d\Sigma^\mu$ the normal
volume element with respect to the metric appearing in \mkg. The
modes are also complete on ${\cal I}^-$
 \eqn\wsz{2i \int_0^\infty
{d\omega \over 2\pi} \bigl( u^{in*}_\omega (t'^{+}) \p_+
u^{in}_{\omega}(t^+)-\p_+u^{in*}_\omega(t^+)
u^{in}_{\omega}(t'^{+})\bigr)=\delta(t^+-t'^{+})~.} We then expand
\eqn\fxs{ \eta(t^+,t^-)= \int_0^\infty d\omega \bigl( a^{in
\dagger}_\omega u^{in*}_\omega(t^+,t^-)+a^{in}_\omega
u^{in}_\omega(t^+,t^-)\bigr)~,} and the quantum vacuum state is
defined by \eqn\type{a^{in}_\omega| 0_{in} \rangle=0~.}

The full solutions of the wave equation can be written
\eqn\slns{u^{in}_\omega(t^+,t^-)={e^{-i\omega\tau^+}\over
\sqrt{2\omega} } +{e^{-i\omega \tau^-}\over \sqrt{2\omega} }~,}
where the Alexandrov coordinates $\tau^\pm(t^+,t^-)$ are given in
\lightcone.  These solutions obey reflecting boundary condition
\rfl\ along the timelike line $\tau^+=\tau^-$. In the $t^\pm$
fermion coordinates the reflecting boundary conditions are at
\eqn\gpa{t^+_R(t^-)= t^-+\ln{2 \mu} -2\ln(1-\lambda e^{t^-})~.} This
trajectory is everywhere timelike but accelerates and reaches
${\cal I}^+$ at the end of the world \eqn\edx{t^-_{end}=-\ln
\lambda~.}
 The behavior of the tachyon vacuum is similar to
the vacuum in the presence of a mirror moving along the trajectory
\gpa. This latter problem is discussed in \bd.

Condition \type\ defines the linearized quantum state but we
are particulary interested in the behavior of outgoing modes on
${\cal I}^+$, where both $t^+ \to \infty$ and $\tau^+\to \infty$.
In this region the outgoing part of the in modes \slns\ is
\eqn\jjh{\eqalign {t^+\to \infty~, ~~~~~~&u^{in}_\omega \to
\bigl[\sqrt{2\over \mu}
\bigl(e^{-t^-}-\lambda \bigr) \bigr]^{i\omega }~,~~~~t^-<t^-_{end}~,\cr
&u^{in}_\omega \to 0~, ~~~~t^->t^-_{end}~,}} while the natural basis
of positive frequency out modes is
\eqn\outmodes{u^{out}_\omega={e^{-i \omega
t^-}\over\sqrt{2\omega}}~.} Since \jjh\ and \outmodes\ are not the
same, an observer near ${\cal I}^+$ moving along a timelike
trajectory of constant $t^+-t^-$ will detect an outgoing particle
flux prior to the end of the world. An observer accelerating along
constant $\tau^+-\tau^-$ will detect nothing.

Before the end of the world the outgoing quantum state can be
expressed as an excitation of the out vacuum. Define the Bogolubov
coefficients
 \eqn\iih{\alpha_{\omega \omega'}=\langle  u^{out}_{\omega'} |u^{in}_{\omega}
\rangle~,~~~~~~~~\beta_{\omega \omega'}=-\langle
u^{out*}_{\omega'}|u^{in}_{\omega} \rangle ~,} and the out creation
operators \eqn\tpl{a^{out\dagger}_\omega= -i\int_{-\infty
}^{t^-_{end}} {dt^- \over 2\pi} u^{out }_\omega(t^-)\ov{ \p_-} \eta
(t^-) ~,} which create plane waves with support prior to
$t^-_{end}$. Then the outgoing quantum state can be written prior
to the end of the world as \eqn\dzzl{ e^{\half \int_0^\infty
d\omega d \omega' a^{out\dagger}_\omega
(\alpha^{-1}\beta)^*_{\omega \omega'} a^{out\dagger}_{\omega'}}
|0_{out}\rangle~,}where $a_\omega^{out}|0_{out}\rangle=0$.

We wish to emphasize the obvious fact that we have not determined the
quantum state on all of ${\cal I^+}$. Although they remain orthonormal, the
 in modes $(u^{in}_\omega,  u^{in*}_\omega)$ in \slns\ do not
evolve to a complete set of modes on ${\cal I}^+$, and in
particular vanish in no-man's land.  Later on we shall interpret
this as the disappearance of spacetime in that region.

 The outgoing energy
density is given (before the world ends)  on ${\cal I}^+$ by the
Schwarzian from Alexandrov to light cone fermion coordinates:
\eqn\cxg{\eqalign{T_{--}(t^-)&=-{1 \over 12}({\p \tau^- \over \p
t^-})^{3/2} {\p^2 \over \p \tau^{-2}}({\p \tau^- \over \p
t^-})^{1/2}\cr &=  -{\lambda e^{t^-}(2-\lambda e^{t^-}) \over
48(1-\lambda e^{t^-})^{2}} ~.}} Note that the energy density \cxg\
is negative and diverges at the critical retarded time \edx. 
 The form of the stress energy \cxg\ is
similar to that at the edge of the Rindler wedge in the Rindler
vacuum, or at the horizon of a black hole in the Boulware
vacuum.\foot{This raises the very interesting question of whether
or not there exists an alternate vacuum which is the analog of the
Hartle-Hawking vacuum for this problem and has a finite stress
energy tensor.} It can be viewed as a Casimir energy arising from
the reflecting boundary condition imposed at the tachyon wall.
This behavior is exactly what one expects viewing the moving
tachyon wall as a moving mirror. In the next subsection we shall
see from the leg pole transform that the spacetime picture is
similar, except that this impenetrable boundary is softened at the string
scale.

In closing this section we note the preceding analysis breaks down
along the reflecting boundary \gpa\ because, among other reasons,
the Casimir energy density is diverging there. The best way to
overcome this problem is to work in the free fermion picture, to
which we turn in section 4.

\subsec{Spacetime particle production}

Now we leg pole transform to the spacetime picture. This is easily
accomplished by simply replacing the expression \jjh\ for the in
modes $u^{in}_\omega (t^-)$ in the out region with their leg pole
transforms $S_\omega(t^-)$ in equation \mode. While the modes in
\jjh\ terminate abruptly at the end of the world, the $S_\omega$
modes decay exponentially after the end of the world. Hence,
spacetime dissolves away on a time period of order the string
time. Nevertheless prior to this cataclysmic event, at the level
of free fields all the information about the incoming quantum
state is transmitted to ${\cal I^+}$.

Viewing the tachyon wall in spacetime as a soft moving mirror, this seems
a natural conclusion.
However, from another point of
view it seems strange.  The modes \slns\ vanish identically in the
region above the null tachyon wall, since the Alexandrov
coordinates simply do not extend into that region. Why can't they tunnel
through the wall with exponentially small amplitudes? Why can't
quantum correlations be lost to the no-man's land after the end of
the world? We shall argue below in section 5 that, while neither
of these phenomena occur in the perturbative collective field
description, they in fact can be seen to occur in the
non-perturbative free fermion formulation.

It is useful to give a quantitative measure of the rate at which
spacetime disappears. To this end, we will consider the
expectation value of the following commutator \eqn\commutS{ \Delta
(q, q';t) = \langle [\partial_t S(t,q), S(t,q')] \rangle~,} where
the expectation value is taken in the state $|0_{in} \rangle$. In
field theory we always have \eqn\dtr{\Delta (q,
q';t)=\delta(q-q')~,} independently of the quantum state. It is
easy to see that, for normalizable perturbations of the static incoming Fermi
sea, \dtr\ remains exactly valid, and the notion of a complete
${\cal I}^+$ remains intact. In fact, if we drain the Fermi sea for
a very long time, but then refill it, \dtr\ is still maintained
everywhere on ${\cal I}^+$. In contrast we shall see
momentarily that tachyon condensation causes $\Delta$ to approach
zero in the far future of ${\cal I}^+$.

 On ${\cal I}^+$,
$\Delta$ has a finite limit and so cannot be a function of
$(t^+,t'^+)$. It may be written in terms of the $S_k$ modes as
\eqn\deltaS{ \Delta(t^-, t'^-) = i\int_0^{\infty} {dk \over \pi}
\bigl(S_k(t^-)
\partial S_k^*(t'^-) - S_k^*(t^-) \partial
S_k(t'^-)\bigr) ~.} Since this integral is hard to deal with, we will
compute this object in a different way: as the double
 leg pole
transform of the $\eta$ commutator. This gives
\eqn\deltaKderive{\eqalign{ \Delta(t^-, t'^-) = &\int dv^-
dv'^-{K\Big( v^- + \ln\big(1 + \lambda e^{t^- -v^-} \big) \Big)
\over 1 + \lambda e^{t^- -v^-} }
 {K\Big( v'^- + \ln\big(1 + \lambda e^{t'^- -v'^-}
\big) \Big) \over 1 + \lambda e^{t'^- -v'^-} } ~\times \cr \times~
& \langle [\partial_t \eta(t^- -v^- + \ln(\sqrt{\mu/2})), 
\eta(t'^- -v'^- + \ln(\sqrt{\mu/2}))] \rangle ~.}}
Since $\eta$ has a standard kinetic term \actiontausigma\ in
$(\tau, \sigma)$ coordinates, we have  \eqn\aaa{ \langle
[\partial_\tau \eta(\tau^-), \eta(\tau'^- )] \rangle = \delta(\tau^- -
\tau'^-) ~.} After a
change of variables, the double integral can be written as
\eqn\deltak{\eqalign{ \Delta(t^-, t'^-) &= \int_{t^- +\ln
\lambda}^\infty  dv^- \int_{t'^- +\ln \lambda}^\infty dv'^-
K(v^-)K(v'^-) \delta(v^--t^- -v'^-+t'^-) (1-\lambda e^{t^- -
v^-})\cr &= \int_{t^- +\ln \lambda}^\infty dv^- ~(1-\lambda e^{t^-
- v^-}) ~K(v^-)K(v^-- t^- + t'^-) ~.}} For both $t^-$ and  $t'^-$
large and negative, the above expression is well approximated by
\eqn\deltaExact {\int_{-\infty}^\infty  dv^- ~K(v^-)K(v^- - t^- +
t'^-) = \delta(t'^- - t^-)~.} Thus, the commutator \commutS\ reduces
for negative $t^-$ to its expected form.  For $t^-$ in no-man's
land, however, it is quite different, approaching zero as
$t^-$ grows, thus signifying the disappearance of spacetime degrees of
freedom.

An informative expression can be given for the integral of
$\Delta$ over $\cal I^+$ with respect to one of its arguments. We
find \eqn\dsz{\int dt'^-\Delta(t'^-,t^-)=\int_{t^-}^\infty
dv^-(1-\lambda e^{t^--v^-})K(v^-)~.} This goes to the field theory
answer $\int\Delta=1$ in the far past, but again approaches zero in
the far future.

The discussion of this subsection employed the collective field
approximation. $\Delta$ can in fact be computed exactly (including
tunneling effects) using the
fermion picture. Surprisingly (at least to us), we will see in the
next section that the results \deltak\ and \dsz\ are exact.

\newsec{The fermion picture}

The above analysis was performed in the collective field
description of the underlying fermion degrees of freedom. In order
to understand the nonperturbative issues, as well as the potential
breakdown of the semiclassical picture at the tachyon wall,
we must go back to the exact
quantum fermion description. A useful reference, whose notation we largely
follow,  is \greg.

Our first task is to write down an exact {\it static} quantum state
which we denote $\mu_R$, in which the Fermi sea is filled up to
the level $-\mu$ on the right hand side of the barrier, but is as
empty as possible on the left side of the barrier. Because of the
exponentially small tunneling, there are no energy eigenstates with no
fermions whatsoever on the left. A complete basis of orthonormal
hamitonian eigenstates obeying \foot{Notice that $\nu$ is
the energy below the top of the potential.}
\eqn\rtip{HW_\nu^{R,L}=i\p_t
W_\nu^{R,L}=-\nu W_\nu^{R,L}}  are provided by the functions
\eqn\dkkk{W_\nu ^R(x,t)=W(\nu,x)e^{i\nu t},~~~~~ W_\nu
^L(x,t)=(W(\nu,-x)+r(\nu) W(\nu,x))e^{i\nu t},} where $r(\nu)$
(whose explicit form shall not be needed) vanishes exponentially
at large  $\nu$ and the Whittaker functions $W(\nu,x)$
obey\greg \eqn\dfy{ W(\nu,x) \sim {1 \over (2 \pi x
\sqrt{1+e^{2\pi\nu}})} (\sqrt{1+e^{2\pi\nu}} - e^{\pi\nu})^{-1/2}
\sin(x^2/4 - \nu \ln x + \Phi(\nu)) } for $x \gg \nu$ and
\eqn\dfy{ W(\nu,x) \sim {1 \over (2 \pi x
\sqrt{1+e^{2\pi\nu}})} (\sqrt{1+e^{2\pi\nu}} - e^{\pi\nu})^{1/2}
\cos(x^2/4 - \nu \ln x + \Phi(\nu)) } for $|x| \gg \nu$ and $x$
negative (for a definition of the phase $\Phi(\nu)$, see \greg). 
Hence, the mode $W^R$ ($W^L$) is supported largely on the
right (left) side of the barrier.

We now expand the Fermi field in terms of creation and
annihilation operators as \eqn\rss{\Psi(x,t)=\int^\mu_{-\infty} d\nu
a^R_\nu W_\nu^R(x,t) + \int_\mu^{\infty} d\nu b^{R\dagger}_\nu
W_\nu^R(x,t)+\int_{-\infty}^\infty d\nu a^L_\nu W_\nu^L(x,t)~.} The
state $|\mu_R\rangle$, corresponding to filling the right side of
the barrier up to energy $-\mu$,  is defined by
\eqn\szaw{a^R_\nu|\mu_R\rangle =b^R_\nu |\mu_R\rangle =a^L_\nu
|\mu_R\rangle=0~.} Later on we shall need the expression for the
eigenvalue density, $\rho(x)$, defined as \eqn\rhodef{ \rho(x)
\equiv \langle \mu_R | \Psi^\dagger \Psi(x)|\mu_R \rangle_ =
\int^{\infty}_\mu d\nu |W (\nu, x)|^2~. } The asymptotic behavior
of $\rho(x)$ is $\rho(x) \sim x$ for large positive $x$ and
$\rho(x) \sim \sin(x^2/4)/(x\ln(x))$ for large negative $x$.

We might also try to write down a coherent
quantum state of the fermion theory corresponding to the classical
motion of the Fermi sea. However, the difference between the
quantum state describing the original static Fermi sea and the
state describing the draining Fermi sea is too large to be
described as a state in the Hilbert space of the original theory
\SeibergBJ. In the language of \SeibergBJ, it involves
non-normalizable modes. Hence, it is associated to a new
Hamiltonian rather than to a semiclassical quantum state in the
theory governed by the old Hamiltonian.

The classical piece is given in fermion language in equation
\moving.  The surface in phase space $(x,p)$ is governed by a flow
induced by a Hamiltonian $(p^2 - x^2)/2$.  Defining new phase
space coordinates $(y, p_y) \equiv (x - \lambda e^t, p - \lambda
e^t)$, we notice that these have the equations of motion $\dot y =
p_y,~ \dot p_y = y$ and that in these coordinates the system is
governed by a Hamiltonian $(p_y^2 - y^2)/2$ -- while the Fermi
surface of interest is given by $(y^2 - p_y^2) = 2 \mu$. We know the
quantum theory corresponding to this classical limit -- it is
simply the theory of free fermions with potential $-y^2/2$, with
all states up to $\mu$ below the top filled. In terms of the
original $(x,p)$ coordinates the time-dependent Hamiltonian is
\eqn\wws{H_\lambda=\half(p^2-x^2) -\lambda e^t (p-x) ~.}

Fermion correlators in the theory defined by \wws\ are easily
computed in terms of the $\lambda=0$ correlators by transforming
to the $(y,p_y)$ phase space.   For example, \eqn\onepoint{
\langle \Psi^\dagger (x,t) \Psi(x',t')\rangle_\lambda \equiv
\langle\Psi^\dagger (y+\lambda e^t ,t) \Psi(y'+\lambda e^{t
'},t')\rangle_\lambda = \langle\Psi^\dagger (y,t)
\Psi(y',t')\rangle~,} where \eqn\eer{y(x,t) = x-\lambda e^t ~.}
$\langle~ \cdot~\rangle$ and $\langle ~\cdot~\rangle_\lambda$
denote expectation values in the theory on the  static
background, and the time-dependent background, respectively.

Of particular interest are correlators of fermion bilinears which
determine the evolution of the shape of the Fermi sea. 
We have \eqn\onepoint{
\langle\Psi^\dagger \Psi(x,t) \Psi^\dagger
\Psi(x',t')\rangle_\lambda \equiv \langle\Psi^\dagger
\Psi(y,t) \Psi^\dagger \Psi(y',t')\rangle~ .} This
determines the two-point function of the collective variable
(eigenvalue density)  \eqn\dazc{ \varphi_0 + 
\sqrt{\pi} \partial_x \eta \sim
\Psi^\dagger \Psi~.} The values of  such correlators on ${\cal
I}^+$ determine the outgoing quantum state on ${\cal I}^+$. It is
easy to see that the exact relation \onepoint\  confirms, in
fermion language, our perturbative computations using the
collective variable $\eta$. That is, the correlators of the previous
section are the semiclassical
limit of the exact correlators. Hence, the conclusions of the previous
section are unaffected by the breakdown of perturbation theory at
the tachyon wall.

To be very concrete about this let us compute the exact expression
for the equal time commutator $\Delta$ in \commutS.  We begin with
\eqn\commutator {\langle [ i\partial_t(\Psi^\dagger \Psi(x,t)),
\Psi^\dagger \Psi(x',t) ] \rangle_\lambda ~.}
 Define
 \eqn\mmk{G(x,x';t) \equiv
\langle \Psi^\dagger(x',t) \Psi(x,t))\rangle_\lambda ~.} Then, it
can be shown using $H\Psi=i\p_t\Psi$ that \eqn\commutatorA
{\langle [ i\partial_t(\Psi^\dagger \Psi(x,t)), \Psi^\dagger
\Psi(x',t) ] \rangle_\lambda = \Re \Big (
\partial_x^2 G(x,x';t) \delta(x-x') - G(x,x';t)\partial_x^2
\delta(x-x') \Big )~. } By integrating this against a test function,
and using the fact that $\Re(G(x,x';t)) = \Re(G(x',x;t))$, we get that
\eqn\commutatorB {\langle [ i\partial_t(\Psi^\dagger \Psi(x,t)),
\Psi^\dagger \Psi(x',t) ] \rangle_\lambda =
- \partial_x(\rho(x,t)\partial_x\delta(x-x'))~.
} The collective field interpretation of this quantity is
\eqn\commutatorB {\langle [ i\partial_t(\Psi^\dagger \Psi(x,t)),
\Psi^\dagger \Psi(x',t) ] \rangle_\lambda =
\partial_x\partial_{x'} \langle [\dot \eta(x,t),
\eta(x',t)]\rangle = \partial_x
\partial_{x'} \delta(x-x')~.} We now can integrate with respect to
both $x$ and $x'$, to obtain a fermionic expression which should
generalize the $\delta$-function: \eqn\commutatorC{ -\int^{x'} dy'
\int^{x} dy\partial_y(\rho(y,t)\partial_y\delta(y-y')) =
\rho(x,t)\delta(x-x') = \rho(x - \lambda e^t) \delta(x-x')~. } This
expression, when leg pole transformed, gives the exact expression
for $\Delta$. In order to compute the leg pole transform, we use
the $t^\pm$ coordinates on ${\cal I}^+$: $2t=t^+ + t^- = 2t' =
t'^+ + t'^-$ and $x = e^{(t^+ - t^-)/2}$, $x' = e^{(t'^+ -
t'^-)/2}$,   for $t^+ \rightarrow \infty$. We get then
\eqn\commutatorC{\rho(x - \lambda e^t)\delta(x-x') =
e^{(t^--t^+ )/2 } \rho\big(
e^{(t^+-t^-)/2} (1 - \lambda e^{t^-})\big) \delta(t'^- - t^-)~.}
The presence of the
$e^{t^+/2}$ factor in the argument of $\rho$ means that only the
asymptotic behavior of $\rho$ will be relevant in ${\cal I}^+$ 
limit, which is \eqn\lpk{\lim_{t^+\to \infty}
e^{-t^+/2}\rho(e^{t^+/2}y)=\Theta(y)~.} 
To take the leg pole transform, we
want to compute \eqn\deltaF{\eqalign{ &\int dv^-
dv'^- K(t^--v^-) K(t'^- - v'^-) e^{(v^--t^+ )/2 } \cr
&~~~~~~~~~~~~~~~~~~~~~~~~ \rho\big( e^{(t^+-v^-)/2}(1 - \lambda
e^{v^-})\big) \delta(v^- -v'^-)  \cr &=\int dv^-
K(t^--v^-) K(t'^- - v^-) e^{(v^--t^+ )/2 } \rho\big(
e^{(t^+-v^-)/2} (1 - \lambda e^{v^-})\big) ~.}}
The integral over  $v^-
>-\ln \lambda$ vanishes in the limit, as can be seen from \lpk.
The integral over the rest of $v^-$ gives simply
\eqn\deltaSAME{\eqalign{\Delta(t^-,t'^-)&= \int_{-\infty}^{-\ln
\lambda} dv^- K(t^--v^-) K(t'^- - v^-) \big(1 - \lambda e^{v^-}
\big)\cr & = \int_{t^-+\ln \lambda}^{\infty} dv^- K(v^-) K(v^- -
t^- + t'^-) \big(1 - \lambda e^{t^--v^-} \big)~,}} which is
obviously the same as \deltak.

Hence, we find that the exact fermion analysis leads to the same
conclusions as the
collective field analysis concerning the outgoing state on ${\cal I}^+$.

\newsec{Nonperturbative effects}

In this section, we will, by example, point out why the tachyon
theory ceases to be unitary when non-perturbative effects are
included. We will use the recently-understood process of D0-brane
creation and decay \refs{\kms,\McGreevyEP} to illustrate the
issues. \foot{The example we give arose in discussion with
J.Polchinski.}

An unstable D0-brane is represented as a single fermion,
corresponding to a single eigenvalue of the matrix model,
 travelling on a classical trajectory. 
When it is sufficiently near the Fermi surface, the
classical fermion can be rewritten as a perturbation of the Fermi
sea, and interpreted as a packet of closed string radiation. Far
away from the sea it must have a different description: that of a
D0-brane on which open strings can end.  Figure 2(a) shows the
spacetime process of D0-brane creation and subsequent decay. 
The same packet of closed string radiation could be travelling on top
of our time-dependent Fermi sea. (In the past, the static and
time-dependent Fermi seas are essentially identical.) It is
possible to choose a set-up in which  the single fermion
is separated from the Fermi sea in the far future because the
Fermi sea drains away before the estranged fermion can reach it.
The spacetime picture of this process is shown in Figure 2(b). The
D0-brane does not decay -- instead, it appears in the no-man's land
behind the wall. These nonperturbative degrees of freedom are free
to explore the entire coordinate region covered by the $t^\pm$
fermion coordinates. This clearly demonstrates that while the
tachyon theory as we have described in the collective picture is
perturbatively unitary, nonperturbatively it cannot be so. It is
missing all degrees of freedom in the no-man's land, and has no
hope of describing them since the Fermi surface does not extend
into that region.

\fig{(a) In a static background, 
closed string radiation creates an unstable D0-brane,
which subsequently decays back into closed strings.
(b) Closed string radiation creates a D0-brane which
does not decay in time to decay into spacetime modes. 
}{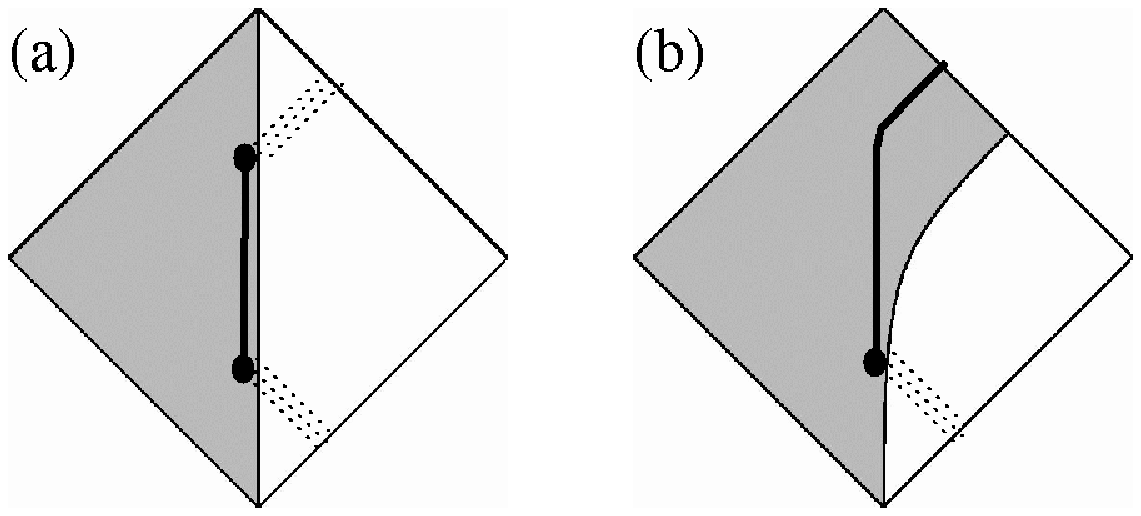}{0.0truein}

Of course this is not a crisis for the matrix model because the
nonperturbative free fermion theory is perfectly unitary.
Nevertheless, there is no unitary spacetime evolution, because
spacetime -- as described by the fluctuating Fermi surface -- does
not extend into the no-man's land. This may be an interesting
lesson for trying to understand black hole physics in string
theory.

\centerline{\bf Acknowledgements}
 This work was supported in part by DOE grant DE-FG02-91ER40654 and the
Harvard Society of Fellows. We are grateful to T. Banks, Wei Li, J.
Maldacena, S.Minwalla, J. Polchinski, N. Seiberg, and T.
Takayanagi for useful conversations.

 \listrefs
\end